\documentclass[12pt]{article}
\usepackage{amsmath} 
\input{epsf} 
\def\ltap{\raisebox{-.6ex}{\rlap{$\,\sim\,$}} \raisebox{.4ex}{$\,<\,$}} 
\def\gtap{\raisebox{-.6ex}{\rlap{$\,\sim\,$}} \raisebox{.4ex}{$\,>\,$}}

\newcommand\as{\alpha_{\mathrm{S}}} 
\newcommand\f[2]{\frac{#1}{#2}}

\def\beq{\begin{equation}} 
\def\eeq{\end{equation}} 
\def\beeq{\begin{eqnarray}} 
\def\eeeq{\end{eqnarray}} 
 
\def\to{\rightarrow} 
\def\nn{\nonumber}

\def\ms{${\overline {\rm MS}}$}

\def\b0{b_0}

\def\tL{{\widetilde L}}

\begin{document} 

\begin{titlepage}
\renewcommand{\thefootnote}{\fnsymbol{footnote}}
\begin{flushright}
    CERN--TH/2003-026   \\ hep-ph/0302104
     \end{flushright}
\par \vspace{10mm}

\begin{center}
{\Large \bf The $q_T$ spectrum of the Higgs boson at the LHC
\\[1.ex]
in QCD perturbation theory}

\end{center}
\par \vspace{8mm}
\begin{center}
{\bf G. Bozzi${}^{(a,b)}$, S. Catani${}^{(b,c)}$, D. de Florian${}^{(d)}$\footnote{Partially supported by Fundaci\'on Antorchas and CONICET.} and M. Grazzini${}^{(a,b,c)}$}\\

\vspace{5mm}

${}^{(a)}$Dipartimento di Fisica, Universit\`a di Firenze, I-50019 Sesto Fiorentino, Florence, Italy\\

${}^{(b)}$INFN, Sezione di Firenze, I-50019 Sesto Fiorentino, Florence, Italy\\

${}^{(c)}$Theory Division, CERN, CH-1211 Geneva 23, Switzerland \\

${}^{(d)}$Departamento de F\'\i sica, FCEYN, Universidad de Buenos Aires,\\
(1428) Pabell\'on 1 Ciudad Universitaria, Capital Federal, Argentina

\vspace{5mm}

\end{center}

\par \vspace{2mm}
\begin{center} {\large \bf Abstract} \end{center}
\begin{quote}
\pretolerance 10000

We consider the transverse-momentum ($q_T$) distribution of 
Higgs bosons produced at hadron colliders.
We use a formalism that 
uniformly treats both the small-$q_T$ and large-$q_T$ regions in
QCD perturbation theory.
At small $q_T$ ($q_T \ll M_H$, $M_H$ being the mass of the Higgs boson),
we implement an all-order resummation of logarithmically-enhanced contributions
up to next-to-next-to-leading logarithmic accuracy. At large $q_T$
($q_T \gtap M_H$), we use fixed-order perturbation theory up to next-to-leading 
order. The resummed and fixed-order approaches are consistently matched by
avoiding double-counting in the intermediate-$q_T$ region. In this region,
the introduction of unjustified higher-order terms is avoided by imposing
unitarity constraints, so that the integral of the $q_T$ spectrum exactly
reproduces the perturbative result for the total cross section up to 
next-to-next-to-leading order. Numerical results at the LHC are presented.
These show that the main features of the $q_T$ distribution are quite stable
with respect to perturbative QCD uncertainties.

\end{quote}

\vspace*{\fill}
\begin{flushleft}
     CERN--TH/2003-026 \\ February 2003

\end{flushleft}
\end{titlepage}

\setcounter{footnote}{1}
\renewcommand{\thefootnote}{\fnsymbol{footnote}}

The Standard Model (SM) of electroweak interactions has been
spectacularly confirmed by experimental data.
However the mechanism of mass generation remains to be understood. 
In its minimal version, the model predicts the existence of a scalar particle,
the Higgs boson [\ref{Gunion:1989we}], 
as a vehicle of electroweak symmetry breaking,
but this particle has so far eluded experimental
discovery. The LEP collaborations  have put a lower limit on 
the mass $M_H$ of the SM Higgs boson at about $114$~GeV [\ref{leplimit}], 
whereas fits to electroweak
data prefer $M_H\ltap 200$~GeV at $95\%$ CL [\ref{lepewfit}].
The next search for Higgs boson(s) will be carried out at hadron colliders,
namely the Fermilab Tevatron [\ref{Carena:2000yx}] 
and the CERN LHC [\ref{atlascms}].

The main SM Higgs production mechanism at hadron colliders is the gluon
fusion process.
At leading order (LO), ${\cal O}(\as^2)$, in the QCD coupling $\as$
this process occurs 
through a heavy-quark (top-quark) loop and,
being a gluon-initiated process, it is expected to receive
large radiative corrections.
It is thus important to perform an accurate evaluation of higher-order
QCD contributions, together with a reliable
estimate of the associated theoretical uncertainty.

The next-to-leading order (NLO) perturbative corrections to the total cross 
section for Higgs boson production via gluon fusion were computed 
in Refs.~[\ref{Dawson:1991zj}]
(in the limit of an infinitely-heavy top quark)
and [\ref{Spira:1995rr}] (including the dependence on the finite mass
$M_t$ of the top quark) and 
were found to be large (of the order of 80--100\%), thus
casting doubts upon the reliability of the perturbative expansion.
In the last two years much effort has been devoted to
improving the accuracy of the perturbative
calculation.
In the large-$M_t$ limit, 
the next-to-next-to-leading order (NNLO) contribution
has been 
computed in Ref.~[\ref{NNLOtotal}]
and still higher-order 
contributions 
have been 
evaluated in Ref.~[\ref{Giele:2002hx}] by implementing soft-gluon resummation.
Since these beyond-NLO corrections
are moderate, the perturbative QCD  predictions for the total
cross section are under good control now.

In this letter we consider a less inclusive observable,
the transverse-momentum ($q_T$) distribution 
of the Higgs boson.
An accurate theoretical prediction of this observable at the LHC
[\ref{atlascms}] can be important to enhance the statistical significance 
of the signal over the background and to improve strategies for the extraction 
of the Higgs boson signal.

When studying the $q_T$ distribution of the Higgs boson in QCD perturbation
theory, it is convenient to start by considering separately
the large-$q_T$ and small-$q_T$ regions.
Roughly speaking, the large-$q_T$ region is identified by
the condition $q_T \gtap M_H$.
In this region, the perturbative series is controlled by a small expansion
parameter, $\as(M_H^2)$, and calculations based on the truncation of the series
at a fixed-order in $\as$ are theoretically justified\footnote{We are not 
considering the extreme limit $q_T \gg M_H$, where a resummation of 
enhanced perturbative terms is required [\ref{Berger:2001wr}].}.
In the small-$q_T$ region ($q_T\ll M_H$), where the bulk of events is produced,
the convergence of the fixed-order expansion is spoiled, since
the coefficients of the perturbative series in $\as(M_H^2)$ are enhanced
by powers of large logarithmic terms, $\ln^m (M_H^2/q_T^2)$. To obtain
reliable perturbative predictions, these terms have 
to be systematically resummed to all orders in $\as$ [\ref{Dokshitzer:hw}].
The fixed-order and resummed approaches have
then
to be consistently matched at intermediate values of $q_T$, 
so as to avoid the introduction of ad-hoc 
boundaries between the large-$q_T$ and small-$q_T$ regions.

Higgs boson production at large $q_T$ has to be accompanied by the radiation
of at least one recoiling parton, so the LO term for this observable is 
of ${\cal O}(\as^3)$.
The LO calculation was
reported
in Ref.~[\ref{Ellis:1987xu}];
it shows that the large-$M_t$
approximation works well as long
as both $M_H$ and $q_T$ are smaller than $M_t$.
Similar results on the validity of the large-$M_t$ approximation
were obtained in the case of the associated production of a Higgs boson plus 
2 jets (2 recoiling partons at large transverse momenta) [\ref{DelDuca:2001fn}].
In the framework of the large-$M_t$ approximation, the NLO QCD corrections
to the transverse-momentum distribution of the Higgs boson
were computed first numerically [\ref{deFlorian:1999zd}]
and later analytically [\ref{Ravindran:2002dc}, \ref{Glosser:2002gm}].
In the large-$q_T$ region, the overall effect of the NLO corrections 
to the $q_T$ distribution is of the
same size as that of the NLO corrections to the total cross section.

The method to systematically perform all-order resummation of
logarithmically-enhanced terms at small $q_T$ is known 
[\ref{Dokshitzer:hw}, \ref{Parisi:1979se}--\ref{Collins:1984kg}]
(see also the list of references in Sect.~5 of
Ref.~[\ref{Catani:2000jh}]). 
To correctly take into account the kinematics constraint of transverse-momentum
conservation, the resummation procedure has to be carried out in
$b$ space, where the impact parameter $b$ is the variable conjugate to 
$q_T$ through a Fourier transformation. In the case of the Higgs boson, 
$b$-space resummation has been explicitly worked out at
leading logarithmic (LL), next-to-leading logarithmic (NLL) 
[\ref{Catani:vd}, \ref{Kauffman:cx}]
and next-to-next-to-leading logarithmic (NNLL) [\ref{deFlorian:2000pr}] level.
The $q_T$ distribution is then obtained by performing the inverse Fourier
(Bessel) transformation with respect to $b$. Various implementation
formalisms [\ref{Collins:1984kg}, \ref{Collins:va}--\ref{Catani:2000vq}]
have been proposed to transform the resummed expressions 
back to $q_T$ space and to perform the matching with the fixed-order results
at large $q_T$. Phenomenological applications to the Higgs boson
$q_T$ distribution have been presented in 
Refs.~[\ref{Hinchliffe:1988ap}, \ref{Kauffman:cx}, \ref{Kauffman:1991jt}--\ref{Berger:2002ut}], 
by combining resummed and fixed-order perturbation theory at different levels
of theoretical accuracy.

In the following we use the formalism described in 
Ref.~[\ref{Catani:2000vq}] to compute  
the Higgs boson $q_T$ distribution at the LHC. In particular, we combine the
most advanced perturbative information that is available at present:
NNLL resummation at small $q_T$ and NLO calculations at large $q_T$.
More details will be given elsewhere.

We consider the collision of two hadrons $h_1$ and $h_2$ with centre-of-mass
energy $\sqrt{s}$.
According to the QCD factorization theorem (see Ref.~[\ref{Collins:gx}] and 
references therein),
the transverse-momentum differential cross section
for the production of the SM Higgs boson can be written as
\begin{equation}
\label{dcross}
\f{d\sigma}{d q_T^2}(q_T,M_H,s)= \sum_{a,b}
\int_0^1 dx_1 \,\int_0^1 dx_2 \,f_{a/h_1}(x_1,\mu_F^2)
\,f_{b/h_2}(x_2,\mu_F^2) \;
\f{d{\hat \sigma}_{ab}}{d q_T^2}(q_T, M_H,{\hat s};
\as(\mu_R^2),\mu_R^2,\mu_F^2) 
\;\;,
\end{equation}
where $f_{a/h}(x,\mu_F^2)$ ($a=q,{\bar q},g$) are the parton densities of 
the colliding hadrons at the factorization scale $\mu_F$,
$d{\hat \sigma}_{ab}/d q_T^2$ are the
partonic cross sections, ${\hat s}=x_1x_2s$ is the partonic centre-of-mass
energy, and $\mu_R$ is the renormalization scale.
Throughout the paper we use parton
densities as defined in the \ms\
factorization scheme, and $\as(q^2)$ is the QCD running coupling in the \ms\
renormalization scheme.

The partonic cross section is computable in QCD perturbation theory and,
as discussed above, it is evaluated by introducing the decomposition
\begin{equation}
\label{resplusfin}
\f{d{\hat \sigma}_{ab}}{d q_T^2}=
\f{d{\hat \sigma}_{ab}^{(\rm res.)}}{d q_T^2}
+\f{d{\hat \sigma}_{ab}^{(\rm fin.)}}{d q_T^2}\, .
\end{equation}
The first term on the right-hand side
contains all 
the logarithmically-enhanced contributions, $\as^n/q_T^2\, \ln^m Q^2/q_T^2$,
at small
$q_T$, and has to be evaluated by resumming them 
to all orders in $\as$.
The second term is free of such contributions, and 
can be computed by fixed-order truncation of the perturbative series.

The resummed component $d{\hat \sigma}_{ac}^{(\rm res.)}$
of the partonic cross section is written as
\begin{equation}
\label{resum}
\f{d{\hat \sigma}_{ac}^{(\rm res.)}}{d q_T^2}(q_T,M_H,{\hat s};
\as(\mu_R^2),\mu_R^2,\mu_F^2) =
\f{1}{2}\int_0^\infty db \; b \;J_0(b q_T) 
\;{\cal W}_{ac}(b,M_H,{\hat s};\as(\mu_R^2),\mu_R^2,\mu_F^2) \;,
\end{equation}
where $J_0(x)$ is the 0th-order Bessel function.
The factor ${\cal W}$ embodies the all-order dependence on 
the large logarithms $L=\ln M_H^2b^2$ at large $b$, which corresponds to the
$q_T$-space terms $\ln M_H^2/q_T^2$ that are 
logarithmically enhanced at small $q_T$ 
(the limit $q_T \ll M_H$ corresponds to $M_Hb \gg 1$, because $b$
is the variable conjugate to $q_T$). Resummation of these large logarithms
is better expressed
by defining the $N$-moments ${\cal W}_N$ of ${\cal W}$
with respect to $z=M^2_H/{\hat s}$ at fixed $M_H$:
\begin{equation}
\label{wndef}
{\cal W}_{ac, \, N}(b,M_H;\as(\mu_R^2),\mu_R^2,\mu_F^2) \equiv
\int_0^1 dz \;z^{N-1} \;{\cal W}_{ac}(b,M_H,{\hat s}=M_H^2/z;
\as(\mu_R^2),\mu_R^2,\mu_F^2) \;.
\end{equation}
The resummation structure of ${\cal W}_{ac, \, N}$ can indeed be organized in 
exponential form as follows:
\begin{align}
\label{wtilde}
{\cal W}_N(b,M_H;\as(\mu_R^2),\mu_R^2,\mu_F^2)
&={\cal H}_N\left(\as(\mu_R^2);M^2_H/\mu^2_R,M^2_H/\mu^2_F\right) \nonumber \\
&\times \exp\{{\cal G}_N(\as(\mu^2_R),bM_H;M^2_H/\mu^2_R,M^2_H/\mu^2_F)\}
\;\;,
\end{align}
where the subscripts denoting the flavour indices are 
understood\footnote{More precisely, we are 
presenting the resummation formulae in a simplified form which is valid when
there is a single species of partons. In general, the exponential
is replaced by an exponential matrix with respect to the flavour indices 
of the partons.}.

All the large logarithmic terms $\as^n L^m=\as^n \ln^m M_H b$
with $1\leq m \leq 2n$ are included (actually, the complete dependence on $b$ 
is included) in the form factor $\exp\{{\cal G}\}$. More importantly,
all the logarithmic contributions to ${\cal G}$ with $n+2 \leq m \leq 2n$
are vanishing. Thus, the exponent ${\cal G}$ can systematically
be expanded as
\begin{align}
\label{gexpan}
{\cal G}_N(\as,bM_H;M^2_H/\mu^2_R,M^2_H/\mu^2_F) &=
\tL \,g^{(1)}(\as \tL)+
g_N^{(2)}(\as \tL;M^2_H/\mu^2_R ) \nn \\
&+
\as \,g_N^{(3)}(\as \tL;M^2_H/\mu^2_R,M^2_H/\mu^2_F)+\dots \;\;,
\end{align}
where $\as=\as(\mu^2_R)$ and the functions $g^{(n)}(\as \tL)$ are defined such
that $g^{(n)}=0$ when $\as \tL=0$. Thus
the term $\tL g^{(1)}$ collects the LL contributions $\as^n \tL^{n+1}$;
the function $g^{(2)}$ resums
the NLL contributions $\as^n \tL^{n}$; $g^{(3)}$ controls the NNLL terms
$\as^n \tL^{n-1}$, and so forth. Note that in the expansion
(\ref{gexpan}) the logarithmic variable $L$ has been replaced by
\begin{equation}
\label{logdef}
\tL=\ln \left(M_H^2b^2/b_0^2+ 1 \right) \;\;,
\end{equation}
where $b_0=2e^{-\gamma}$. In the resummation region $M_Hb \gg 1$,
the replacement is fully legitimate since $\tL \sim L$.
The reason for using $\tL$ rather than $L$ is discussed below.

The function ${\cal H}_N$ in Eq.~(\ref{wtilde}) does not depend on $b$ and,
hence, its evaluation does not require resummation of large logarithmic terms.
It can be expanded in powers of $\as=\as(\mu^2_R)$ as
\begin{align}
\label{hexpan}
{\cal H}_N(\as;M^2_H/\mu^2_R,M^2_H/\mu^2_F)&=\sigma_0 \;\as^2 
\Bigl[ 1+ \f{\as}{2\pi} \,{\cal H}_N^{(1)}(M^2_H/\mu^2_R,M^2_H/\mu^2_F) 
\Bigr. \nn \\
&+ \Bigl.
\left(\f{\as}{2\pi}\right)^2 \,{\cal H}_N^{(2)}(M^2_H/\mu^2_R,M^2_H/\mu^2_F) 
+ \dots\Bigr] \;\;,
\end{align}
where $\sigma_0= G_F/(288 \pi {\sqrt 2})$ is the Born level cross section 
in the large-$M_t$ approximation, and 
$G_F=1.16639\times 10^{-5} \,{\rm GeV}^{-2}$ is the Fermi constant.

The `finite' component $d{\hat \sigma}_{ab}^{(\rm fin.)}$ of the partonic
cross section does not require resummation of large logarithmic terms either.
We compute it as follows: 
\begin{equation}
\label{resfin}
\f{d{\hat \sigma}_{ab}^{(\rm fin.)}}{d q_T^2} =
\Bigl[\f{d{\hat \sigma}_{ab}^{}}{d q_T^2}\Bigr]_{\rm f.o.}
- \Bigl[ \f{d{\hat \sigma}_{ab}^{(\rm res.)}}{d q_T^2}\Bigr]_{\rm f.o.} \;.
\end{equation}
The first term on the right-hand side is the usual perturbative series
for the partonic cross section truncated at a given fixed order in $\as$.
The second term is obtained by truncating the resummed component
in Eq.~(\ref{resum}) at the {\em same} fixed order in $\as$.
The (small-$q_T$) resummed and (large-$q_T$)
fixed-order approaches are thus consistently matched by
avoiding double-counting in the intermediate-$q_T$ region.
This procedure guarantees that the right-hand side of Eq.~(\ref{resplusfin})
contains the full information of the perturbative calculation up to
the fixed order specified by Eq.~(\ref{resum}) plus resummation of
logarithmically-enhanced contributions from higher orders.

A few distinctive features of the formalism described so far
require some comments.

We implement perturbative QCD resummation at the level of the partonic cross
section. In the factorization formula (\ref{dcross}), the parton densities are 
thus evaluated at the factorization scale $\mu_F$, as in the customary
perturbative calculations at large $q_T$. The central value of $\mu_F$ 
and $\mu_R$ has to be set equal to $M_H$, the typical hard scale of the 
process, and the theoretical accuracy of the resummed calculation can be 
investigated as in fixed-order calculations, by varying $\mu_F$ and $\mu_R$
around this central value.

The variables $L$ and $\tL$ are equivalent to organize the resummation 
formalism in the region $M_Hb \gg 1$. The use of the variable $\tL$
is inspired by the procedure introduced in Ref.~[\ref{Catani:1992ua}]
to deal with kinematical constraints when performing
soft-gluon resummation in $e^+e^-$ event shapes. When $M_Hb \ll 1$, 
$\tL \to 0$ and $\exp \{{\cal G}\} \to 1$. Therefore, using the definition 
in Eq.~(\ref{logdef}), we avoid the introduction of all-order contributions 
in the small-$b$ region, where the use of 
the large-$b$ resummation formalism is not justified. 
In particular, $\exp \{{\cal G}\} = 1$ at $b=0$. This implies that the integral
over $q_T$ of $d\sigma/dq_T$ exactly reproduces the fixed-order calculation
of the total cross section. Note that the bulk of the $q_T$ distribution is in
the region $q_T \ltap M_H$. Since resummed and fixed-order perturbation theory
controls the small-$q_T$ and large-$q_T$ regions respectively,
the total cross section constraint mainly acts on the size of the higher-order
contributions introduced in the intermediate-$q_T$ region by the matching
procedure.

It is known
[\ref{Collins:va}--\ref{Ellis:1997ii}, \ref{Guffanti:2000ep}, \ref{Qiu:2000ga}]
that non-perturbative effects have an increasing role in the
$q_T$ distribution as $q_T$ decreases. However, we do not include 
non-perturbative contributions. The main goal of the quantitative study
presented below is to investigate the predictivity of QCD within a purely
perturbative framework. In particular, we want to examine how the 
Higgs boson $q_T$ distribution is affected by perturbative QCD uncertainties,
such as its dependence on scale variations and on higher-order contributions. 

The functions $g_N^{(k)}(\as \tL)$ and the coefficients ${\cal H}_N^{(k)}$
in Eqs.~(\ref{gexpan}) and (\ref{hexpan}) can
be expressed (see for instance Ref.~[\ref{Frixione:1998dw}])
in terms of perturbative coefficients known as
$A^{(n)}$, $B^{(n)}$, $C^{(n)}$ [\ref{Collins:1984kg}]
and $H^{(n)}$ [\ref{Catani:2000vq}].
In particular, $g^{(1)}$ depends on $A^{(1)}$, $g_N^{(2)}$ also depends
on $B^{(1)}$ and $A^{(2)}$ [\ref{Catani:vd}], $g_N^{(3)}$ also depends
on $H^{(1)}$, $C^{(1)}$ [\ref{Kauffman:cx}], 
$B^{(2)}$ [\ref{deFlorian:2000pr},\ref{Glosser:2002gm}]  
and $A^{(3)}$, ${\cal H}_N^{(1)}$ depends on $H^{(1)}$ and
$C^{(1)}$, ${\cal H}_N^{(2)}$ also depends on $H^{(2)}$ and $C^{(2)}$.
We also observe that the functions $g_N^{(2)}$ and $g_N^{(3)}$
receive additional contributions respectively
from the LO and NLO anomalous dimensions
that control the evolution of the parton densities.
The NNLL coefficient $A^{(3)}$ is not yet known.
In the following we assume that
its value is the same as the one [\ref{Vogt:2000ci}]
that appears in resummed calculations of soft-gluon contributions near
threshold. 
The coefficient ${\cal H}_N^{(2)}$ is not known in analytic form either.
However, within our formalism we can exploit the property that the integral
of the $q_T$ distribution exactly matches the fixed-order calculation of the
total cross section. From the known NNLO result for the total cross section
[\ref{NNLOtotal}], we thus extract ${\cal H}_N^{(2)}$ in (approximate)
numerical form.
As pointed out in Ref.~[\ref{Catani:2000vq}], the coefficients
$B^{(n)}$, $C^{(n)}$ and $H^{(n)}$ cannot separately be defined 
without fixing a resummation scheme. Note, however, that the dependence on 
the choice of the resummation scheme cancels by recasting the resummed 
formulae in the form of Eq.~(\ref{wtilde}): 
the functions $g_N^{(k)}(\as \tL)$ and
the coefficients  ${\cal H}_N^{(k)}$in Eqs.~(\ref{gexpan})
and (\ref{hexpan}) are explicitly resummation-scheme independent.

The functions $g_N^{(k)}(\as \tL)$ are singular when $\lambda=\beta_0 \as \tL
\to 1$ ($\beta_0$ is the first coefficient of the QCD $\beta$-function).
The singular behaviour is related to the presence of the Landau pole
in the perturbative running of the QCD coupling $\as(q^2)$.
To properly
define the $b$ integration in Eq.~(\ref{resum}), a prescription 
to deal with these
singularities has to be introduced.
Here we follow Ref.~[\ref{Laenen:2000de}] and deform the integration contour
in the complex $b$ space, as an extension of the
minimal prescription of Ref.~[\ref{Catani:1996yz}].
 
In the following we 
present quantitative results at NLL+LO and NNLL+NLO 
accuracy. We implement Eqs.~(\ref{resplusfin}) and (\ref{resfin}).
At NLL+LO accuracy, we compute $d\sigma^{(\rm res.)}$ at NLL accuracy 
(we include the coefficient ${\cal H}_N^{(1)}$
and the functions $g_N^{(1)}$ and $g_N^{(2)}$), and we match it with 
$[ d\sigma ]_{\rm f.o.}$ evaluated at LO (i.e. at ${\cal O}(\as^3)$).
At NNLL+NLO accuracy, we also include ${\cal H}_N^{(2)}$ and $g_N^{(3)}$
in the resummed component and we evaluate $[ d\sigma ]_{\rm f.o.}$ 
at NLO (i.e. at ${\cal O}(\as^4)$). As for the evaluation of 
$[ d\sigma ]_{\rm f.o.}$, we use the Monte Carlo program 
of Ref.~[\ref{deFlorian:1999zd}].
The numerical results are obtained by using the MRST2001 set of
parton distributions [\ref{Martin:2001es}] and choosing
$M_H=125$~GeV. At NLL+LO we use LO parton densities and
1-loop $\as$, whereas at NNLL+NLO we use NLO parton densities 
and 2-loop $\as$.

\begin{figure}[htb]
\begin{center}
\begin{tabular}{c}
\epsfxsize=12truecm
\epsffile{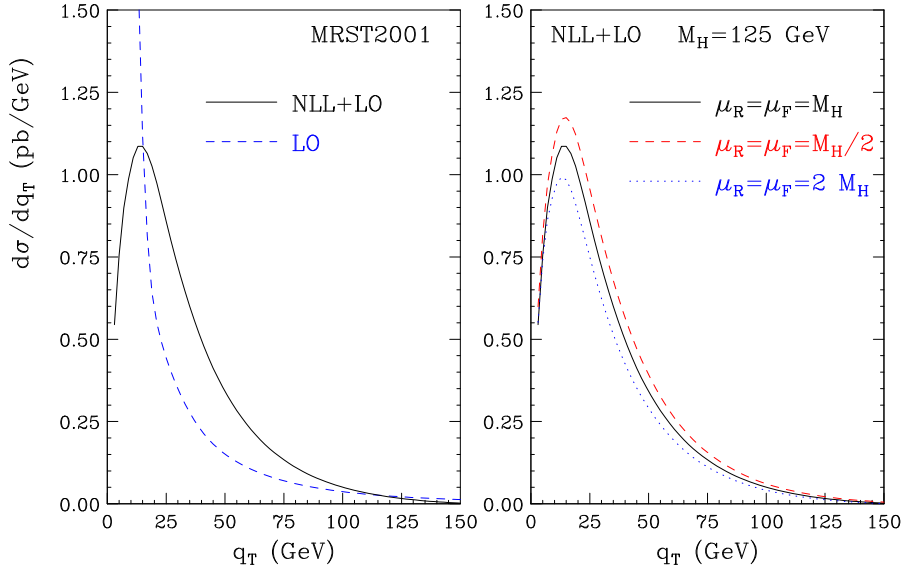}\\
\end{tabular}
\end{center}
\caption{\label{fig1}
{\em 
LHC results at NLL+LO accuracy.}}
\end{figure}

The NLL+LO results at the LHC are shown in Fig.~\ref{fig1}.
In the left-hand side, the full NLL+LO result (solid line)
is compared with the LO one (dashed line)
at the default scales $\mu_F=\mu_R=M_H$.
We see that the LO calculation diverges to $+\infty$ as $q_T\to 0$. 
The effect of the resummation is relevant below $q_T\sim 100$~GeV.
In the right-hand side we show the NLL+LO band that is obtained
by varying $\mu_F=\mu_R$ between $1/2 M_H$ and $2M_H$.
The scale dependence increases from about $\pm 10\%$ at the peak
to about $\pm 20\%$ at $q_T=100$~GeV.
The integral of the resummed curve
is in good agreement
with the value of the NLO 
total cross section evaluated with
LO parton densities and 1-loop $\as$, the small difference being due to
the (improvable) numerical precision of our code.

\begin{figure}[htb]
\begin{center}
\begin{tabular}{c}
\epsfxsize=12truecm
\epsffile{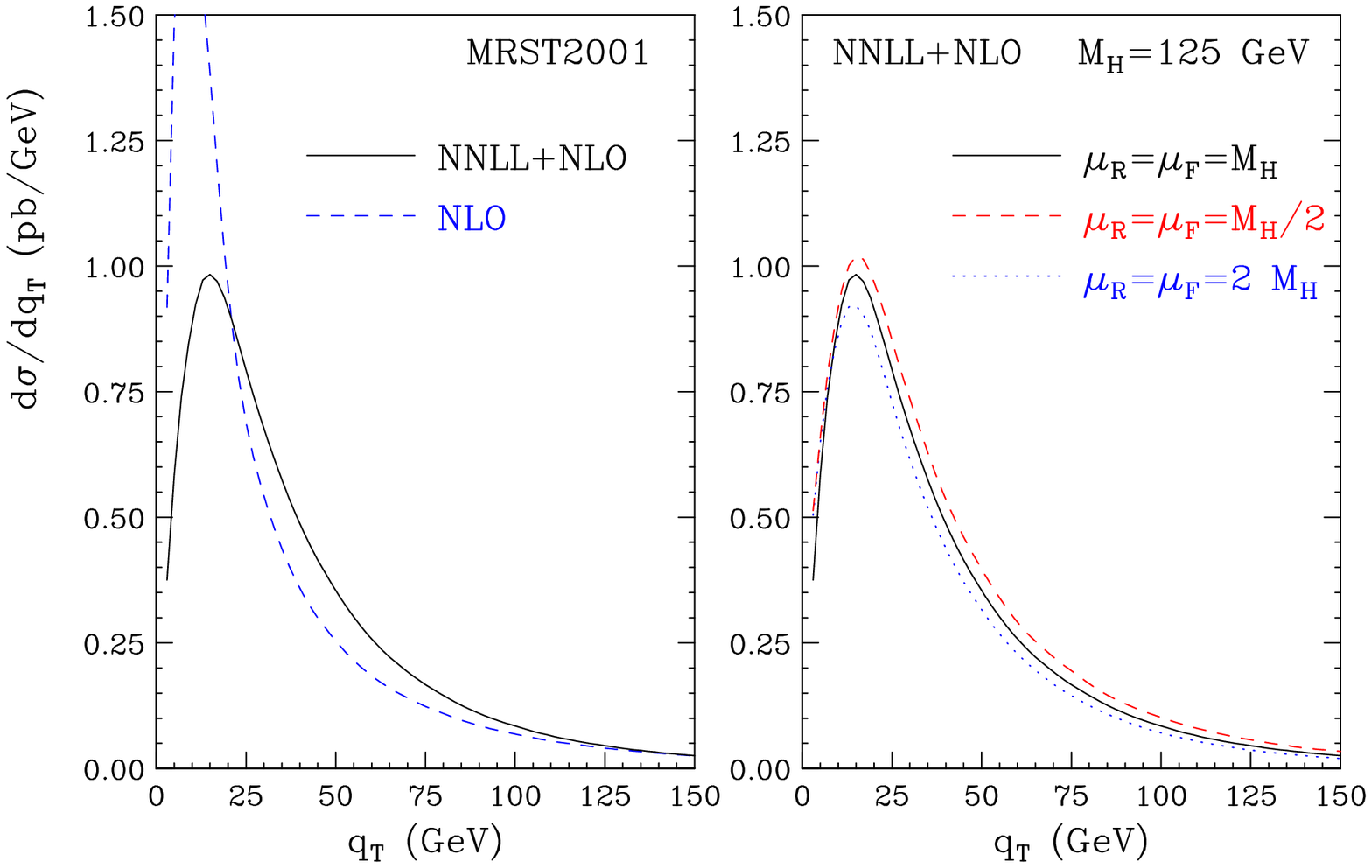}\\
\end{tabular}
\end{center}
\caption{\label{fig2}
{\em 
LHC results at NNLL+NLO accuracy. }}
\end{figure}

The NNLL+NLO results at the LHC are shown in Fig.~\ref{fig2}.
In the left-hand side, the full result (solid line)
is compared with the NLO one (dashed line) at the
default scales $\mu_F=\mu_R=M_H$.
The NLO result diverges to $-\infty$ as $q_T\to 0$ and, at small values of 
$q_T$, it has an unphysical peak (the top of the peak is above the vertical
scale of the plot) which is produced by the numerical compensation of negative
leading logarithmic and positive subleading logarithmic contributions.
It is interesting to compare the LO and NLL+LO curves in Fig.~\ref{fig1}
and the NLO curve in Fig.~\ref{fig2}. At $q_T \sim 50$~GeV, the 
$q_T$ distribution sizeably increases when going from LO to NLO and from NLO
to NLL+LO. This implies that in the intermediate-$q_T$ region there are
important contributions that have to be resummed to all orders rather than
simply evaluated at the next perturbative order.
The $q_T$ distribution is (moderately) harder
at NNLL+NLO than at NLL+LO accuracy.
The height of the NNLL peak is a bit lower
than the NLL one. This is mainly due to the fact that
the total NNLO cross section
(computed with NLO parton densities and 2-loop $\as$), 
which fixes the value of
the $q_T$ integral of our resummed result,
is slightly smaller than
the NLO one, whereas the high-$q_T$ tail is higher at NNLL order,
thus leading to a reduction of the cross section at small $q_T$.
We find that the contribution of $A^{(3)}$ (recall that we are using 
an educated guess
on the value of the coefficient $A^{(3)}$)
can safely be neglected.
The coefficient ${\cal H}_N^{(2)}$ contributes significantly, and enhances
the $q_T$ distribution by roughly $20\%$ in the region of intermediate and
small values of $q_T$.
The resummation effect starts to be
visible
below $q_T\sim 100$~GeV, and 
it increases the NLO result by about $40\%$ at $q_T=50$~GeV.
The right-hand side of Fig.~\ref{fig2} shows the scale dependence computed as
in Fig.~\ref{fig1}. The scale dependence is now about $\pm 6\%$ at the peak
and increases to $\pm 20\%$ at $q_T=100$~GeV.
Comparing Figs.~1 and 2, we see that the NNLL+NLO band is smaller 
than the NLL+LO one and overlaps with the latter at $q_T \ltap 100$~GeV.
This suggests a good convergence of the resummed perturbative expansion. 

We have considered perturbative QCD predictions for the Higgs boson  
$q_T$ distribution at the LHC. We have shown that the main features of the
$q_T$ distribution are quite stable with respect to perturbative 
uncertainties (scale variations, inclusion of higher orders in the 
resummed expansion).
More details about the formalism and our numerical results will be presented
in a future publication, where we shall also consider the inclusion of
non-perturbative contributions.
Available studies [\ref{Balazs:2000wv}--\ref{Berger:2002ut}]
of non-perturbative contributions at the LHC 
estimate effects (at most) of the order of a few per cent when 
$q_T \gtap 10$~GeV. These effects are smaller than the resummation effects
examined here.

{\bf  Acknowledgements}

We would like to thank Werner Vogelsang for discussions. 
S.C. wishes to thank Luca Trentadue for their early collaboration 
on Higgs boson production in hadronic collisions.

\section*{References}

\begin{enumerate}

\item \label{Gunion:1989we}
For a review on Higgs physics in and beyond the Standard Model, see
J.~F.~Gunion, H.~E.~Haber, G.~L.~Kane and S.~Dawson,
{\it The Higgs Hunter's Guide} (Addison-Wesley, Reading, Mass., 1990);
M.~Carena and H.~E.~Haber,
preprint FERMILAB-PUB-02-114-T [hep-ph/0208209].

\item \label{leplimit} 

A.~Heister {\it et al.}  [ALEPH Collaboration],
Phys.\ Lett.\ B {\bf 526} (2002) 191;
P.~Abreu {\it et al.}  [DELPHI Collaboration],
Phys.\ Lett.\ B {\bf 499} (2001) 23;
P.~Achard {\it et al.}  [L3 Collaboration],
Phys.\ Lett.\ B {\bf 517} (2001) 319;
G.~Abbiendi {\it et al.}  [OPAL Collaboration],
hep-ex/0209078.

\item \label{lepewfit}
The LEP Collaborations, the LEP Electroweak Working Group and 
the SLD Heavy Flavour Group, 
report LEPEWWG/2002--02, hep-ex/0212036.

\item \label{Carena:2000yx}
M.~Carena {\it et al.},
{\it Report of the Tevatron Higgs working group},
hep-ph/0010338.

\item \label{atlascms}
CMS Coll., {\it Technical Proposal}, report CERN/LHCC/94-38 (1994);
ATLAS Coll., {\it ATLAS Detector and Physics Performance: Technical Design
Report}, Vol. 2, report CERN/LHCC/99-15 (1999).

\item \label{Dawson:1991zj}
S.~Dawson,
Nucl.\ Phys.\ B {\bf 359} (1991) 283;
A.~Djouadi, M.~Spira and P.~M.~Zerwas,
Phys.\ Lett.\ B {\bf 264} (1991) 440.

\item \label{Spira:1995rr}
M.~Spira, A.~Djouadi, D.~Graudenz and P.~M.~Zerwas,
Nucl.\ Phys.\ B {\bf 453} (1995) 17.

\item \label{NNLOtotal}
S.~Catani, D.~de Florian and M.~Grazzini,
JHEP {\bf 0105} (2001) 025;
R.~V.~Harlander and W.~B.~Kilgore,
Phys.\ Rev.\ D {\bf 64} (2001) 013015,
Phys.\ Rev.\ Lett.\  {\bf 88} (2002) 201801;
C.~Anastasiou and K.~Melnikov,
Nucl.\ Phys.\ B {\bf 646} (2002) 220.

\item \label{Giele:2002hx}
S.~Catani, D.~de Florian, M.~Grazzini and P.~Nason,
in  W.~Giele {\it et al.},
hep-ph/0204316, p.~51, 
contributed to the Les Houches 2001
Workshop on {\em Physics at TeV Colliders}.

\item \label{Berger:2001wr}
E.~L.~Berger, J.~w.~Qiu and X.~f.~Zhang,
Phys.\ Rev.\ D {\bf 65} (2002) 034006.

\item \label{Dokshitzer:hw}
Y.~L.~Dokshitzer, D.~Diakonov and S.~I.~Troian,
Phys.\ Rep.\  {\bf 58} (1980) 269.

\item \label{Ellis:1987xu}
R.~K.~Ellis, I.~Hinchliffe, M.~Soldate and J.~J.~van der Bij,
Nucl.\ Phys.\ B {\bf 297} (1988) 221;
U.~Baur and E.~W.~Glover,
Nucl.\ Phys.\ B {\bf 339} (1990) 38.

\item \label{DelDuca:2001fn}
V.~Del Duca, W.~Kilgore, C.~Oleari, C.~Schmidt and D.~Zeppenfeld,
Nucl.\ Phys.\ B {\bf 616} (2001) 367.

\item \label{deFlorian:1999zd}
D.~de Florian, M.~Grazzini and Z.~Kunszt,
Phys.\ Rev.\ Lett.\  {\bf 82} (1999) 5209.

\item \label{Ravindran:2002dc}
V.~Ravindran, J.~Smith and W.~L.~Van Neerven,
Nucl.\ Phys.\ B {\bf 634} (2002) 247.

\item \label{Glosser:2002gm}
C.~J.~Glosser and C.~R.~Schmidt,
JHEP {\bf 0212} (2002) 016.

\item \label{Parisi:1979se}
G.~Parisi and R.~Petronzio,
Nucl.\ Phys.\ B {\bf 154} (1979) 427.

\item \label{Curci:1979bg}
G.~Curci, M.~Greco and Y.~Srivastava,
Nucl.\ Phys.\ B {\bf 159} (1979) 451.

\item \label{Collins:1981uk}
J.~C.~Collins and D.~E.~Soper,
Nucl.\ Phys.\ B {\bf 193} (1981) 381
[Erratum-ibid.\ B {\bf 213} (1983) 545].

\item \label{Kodaira:1981nh}
J.~Kodaira and L.~Trentadue,
Phys.\ Lett.\ B {\bf 112} (1982) 66,
report SLAC-PUB-2934 (1982).

\item \label{Collins:1984kg}
J.~C.~Collins, D.~E.~Soper and G.~Sterman,
Nucl.\ Phys.\ B {\bf 250} (1985) 199.

\item \label{Catani:2000jh}
S.~Catani et al.,
hep-ph/0005025, in the Proceedings of the CERN Workshop on {\it Standard Model
Physics (and more) at the LHC}, eds. G.~Altarelli and M.L.~Mangano
(CERN 2000-04, Geneva, 2000), p.~1.

\item \label{Catani:vd}
S.~Catani, E.~D'Emilio and L.~Trentadue,
Phys.\ Lett.\ B {\bf 211} (1988) 335.

\item \label{Kauffman:cx}
R.~P.~Kauffman,
Phys.\ Rev.\ D {\bf 45} (1992) 1512.

\item \label{deFlorian:2000pr}
D.~de Florian and M.~Grazzini,
Phys.\ Rev.\ Lett.\ {\bf 85} (2000) 4678,
Nucl.\ Phys.\ B {\bf 616} (2001) 247.

\item \label{Collins:va}
J.~C.~Collins and D.~E.~Soper,
Nucl.\ Phys.\ B {\bf 197} (1982) 446.

\item \label{Altarelli:1984pt}
G.~Altarelli, R.~K.~Ellis, M.~Greco and G.~Martinelli,
Nucl.\ Phys.\ B {\bf 246} (1984) 12.

\item \label{Ellis:1997ii}
R.~K.~Ellis and S.~Veseli,
Nucl.\ Phys.\ B {\bf 511} (1998) 649.

\item \label{Frixione:1998dw}
S.~Frixione, P.~Nason and G.~Ridolfi,
Nucl.\ Phys.\ B {\bf 542} (1999) 311.

\item \label{Kulesza:1999gm}
A.~Kulesza and W.~J.~Stirling,
Nucl.\ Phys.\ B {\bf 555} (1999) 279,
JHEP {\bf 0001} (2000) 016.

\item \label{Catani:2000vq}
S.~Catani, D.~de Florian and M.~Grazzini,
Nucl.\ Phys.\ B {\bf 596} (2001) 299.

\item \label{Hinchliffe:1988ap}
I.~Hinchliffe and S.~F.~Novaes,
Phys.\ Rev.\ D {\bf 38} (1988) 3475.

\item \label{Kauffman:1991jt}
R.~P.~Kauffman,
Phys.\ Rev.\ D {\bf 44} (1991) 1415.

\item \label{Yuan:1991we}
C.~P.~Yuan,
Phys.\ Lett.\ B {\bf 283} (1992) 395.

\item \label{Balazs:2000wv}
C.~Balazs and C.~P.~Yuan,
Phys.\ Lett.\ B {\bf 478} (2000) 192.

\item \label{Balazs:2000sz}
C.~Balazs, J.~Huston and I.~Puljak,
Phys.\ Rev.\ D {\bf 63} (2001) 014021;
see also Sect.~5.4 
in  W.~Giele {\it et al.},
hep-ph/0204316,  
contributed to the Les Houches 2001
Workshop on {\em Physics at TeV Colliders}.

\item \label{Berger:2002ut}
E.~L.~Berger and J.~w.~Qiu,
preprint ANL-HEP-PR-02-057 [hep-ph/0210135].

\item \label{Collins:gx}
J.~C.~Collins, D.~E.~Soper and G.~Sterman,
in {\it Perturbative Quantum Chromodynamics},
ed. A.H.~Mueller (World Scientific, Singapore, 1989), p.~1. 

\item \label{Catani:1992ua}
S.~Catani, L.~Trentadue, G.~Turnock and B.~R.~Webber,
Nucl.\ Phys.\ B {\bf 407} (1993) 3.

\item \label{Guffanti:2000ep}
A.~Guffanti and G.~E.~Smye,
JHEP {\bf 0010} (2000) 025.

\item \label{Qiu:2000ga}
J.~w.~Qiu and X.~f.~Zhang,
Phys.\ Rev.\ Lett.\  {\bf 86} (2001) 2724,
Phys.\ Rev.\ D {\bf 63} (2001) 114011.

\item \label{Vogt:2000ci}
A.~Vogt,
Phys.\ Lett.\ B {\bf 497} (2001) 228;
C.~F.~Berger,
Phys.\ Rev.\ D {\bf 66} (2002) 116002.

\item \label{Laenen:2000de}
E.~Laenen, G.~Sterman and W.~Vogelsang,
Phys.\ Rev.\ Lett.\  {\bf 84} (2000) 4296;
A.~Kulesza, G.~Sterman and W.~Vogelsang,
Phys.\ Rev.\ D {\bf 66} (2002) 014011.

\item \label{Catani:1996yz}
S.~Catani, M.~L.~Mangano, P.~Nason and L.~Trentadue,
Nucl.\ Phys.\ B {\bf 478} (1996) 273.

\item \label{Martin:2001es}
A.~D.~Martin, R.~G.~Roberts, W.~J.~Stirling and R.~S.~Thorne,
Eur.\ Phys.\ J.\ C {\bf 23} (2002) 73.

\end{enumerate}

\end{document}